\documentclass[preprint,12pt]{article}

\usepackage{amssymb}
\usepackage{graphicx}
\usepackage{amsmath}
\usepackage{color}

\begin{document}


{\Large \bf Synaptic Plasticity and Spike Synchronisation in Neuronal Networks}
\\
\\
{Rafael R. Borges$^1$, Fernando S. Borges$^2$, Ewandson L. Lameu$^3$, Paulo R. Protachevicz$^4$, Kelly C. Iarosz$^{2,5}$, Iber\^e L. Caldas$^2$, Ricardo L. Viana$^6$, Elbert E. N. Macau$^3$, Murilo S. Baptista$^5$, Celso Grebogi$^5$, Antonio M. Batista$^{2,4,5,7}$}
\\
\\
\\ \noindent{$^1$Department of Mathematics, Federal University of Technology Paran\'a, Apucarana, PR, Brazil.}
\\{$^2$Institute of Physics, University of S\~ao Paulo, S\~ao Paulo, SP, Brazil.}
\\{$^3$National Institute for Space Research, S\~ao Jos\'e dos Campos, SP, Brazil.}
\\{$^4$Post-Graduation in Science, State University of Ponta Gros\-sa, Ponta Grossa, PR, Brazil.}
\\{$^5$Institute for Complex Systems and Mathematical Biology, King's College, University of Aberdeen, Aberdeen, AB24 3UE, UK.}
\\{$^6$Department of Physics, Federal University of Paran\'a, Curitiba, PR, Brazil.}
\\{$^7$Department of Mathematics and Statistics, State University of Ponta Grossa, Ponta Grossa, PR, Brazil.}

\begin{abstract}
Brain plasticity, also known as neuroplasticity, is a fundamental mechanism of neuronal adaptation in response to changes in the environment or due to brain injury. In this review, we show our results about the effects of synaptic plasticity on neuronal networks composed by Hodgkin-Huxley neurons. We show that the final topology of the evolved network depends crucially on the ratio between the strengths of the inhibitory and excitatory synapses. Excitation of the same order of inhibition revels an evolved network that presents the rich-club phenomenon, well known to exist in the brain. For initial networks with considerably larger inhibitory strengths, we observe the emergence of a complex evolved topology, where neurons sparsely connected to other neurons, also a typical topology of the brain. The presence of noise enhances the strength of both types of synapses, but if the initial network has synapses of both natures with similar strengths. Finally, we show how the synchronous behaviour of the evolved network will reflect its evolved topology. 

\end{abstract}


\section{Introduction}

The brain \footnote{The Brain is wider than the Sky,\\ For, put them side by side,\\ The one the other will include\\ With ease, and you beside.\\ Emily Dickinson, Complete Poems. 1924 (1830-1886).} is the most complex organ 
in the human body. It contains approximately $10^2$ billion neurons and $10^3$ 
trillion synaptic connections, where each neuron can be connected to up to 
$10^4$ other neurons \cite{gerstner02}. The neuron is the basic working unit of
the brain and it is responsible for carrying out the communication and the 
processing of information within the brain \cite{sporns05}. Those tasks are 
achieved through neuronal firing spatio-tem\-poral patterns that are depended 
on the neuron own dynamics and the way they are networked.

Towards the goal to understand the brain, over the past several years, 
mathematical models have been introduced to emulate neuronal firing patterns. A
simple model that has been considered to describe neuronal spiking is based on 
the cellular automaton \cite{viana14,borges15}. This model uses discrete state 
variables, coordinates and time \cite{wolfram83}. Another proposed bursting 
behaviour mo\-del is a simplification of the neuron model described by 
differential equations, where the state variables are continuous, while the 
coordinates and the time are discrete \cite{batista12,lameu16a,lameu16b}. 
Recently, Girardi-Schappo et al. \cite{schappo17} proposed a map that 
reproduces neuronal excitatory and autonomous behaviour that are observed 
experimentally.

Differential equations have also been used to model neuronal patterns
\cite{abbott99,batista14,baptista10}. The integrate-and-fire model was
developed by Lapicque in 1907 \cite{lapicque07} and it is still widely used.
But one of the most successful and cerebrated mathematical models using
differential equations was proposed by Hodgkin and Huxley in $1952$
\cite{hodgkin52}. The Hodgkin-Huxley model explains the ionic mechanisms
related to propagation and initiation of action potentials, i.e., the
characteristic potential pulse that propagates in the neurons. In $1984$,
Hindmarsh and Rose \cite{hindmarsh84} developed a model that simulates bursts
of spikes. The phenomenological Hindmarsh-Rose model may be seen as a
simplification of the Hodgkin-Huxley model.

Hodgkin-Huxley neuron networks have been successfully used as a mathematical
model to describe processes occurring in the brain. An important brain activity
phenomenon is the neuronal synchronisation. This phenomenon is related to
cognitive functions, memory processes, perceptual and motor skills, and
information transfer 
\cite{baptista10,baptista08,baptista08plos,cris15,uhlhaas09}.

There has been much work on neuronal synchronisation. Temporal synchronisation
of neuronal activity happens when neurons are excited synchronously, namely
assemblies of neurons fire simultaneously \cite{baptista10,melloni07}. Newly,
Borges and collaborators \cite{borges17} modelled spiking and bursting 
synchronous behaviour in a neuronal network. They showed that not only 
synchronisation, but also the kind of synchronous behaviour depends on the 
coupling strength and neuronal network connectivity. Studies showed that phase 
synchronisation is related to information transfer between brain areas at 
different frequency bands \cite{fell11}. Neuronal synchronisation can be 
related to brain disorders, such as epilepsy and Parkinson's disease. 
Parkinson's disease is associated with synchronised oscillatory activity in 
some specific part of the brain \cite{rubchinsky12}. Based on that, Lameu et 
al. \cite{lameu16} proposed interventions in neuronal networks to provide a 
procedure to suppress pathological rhythms associated with forms of 
synchronisation.

In this review, we focus the attention on the weakly and strongly synchronous
states in dependence with brain plasticity. Brain plasticity, also known as
neuroplasticity, is a fundamental mechanism for neuronal adaptation in response
to changes in the environment or to new situations \cite{benett64}. In 1890,
James \cite{james90} proposed that the interconnection among the neurons in the
brain and so the functional behaviour carried on by neurons are not static.
Experimental evidence of plasticity was demonstrated by Lashley in 1923
\cite{lashley23} through experiments on monkeys. Scientific evidence of
anatomical brain plasticity was published in 1964 by Bennett et al.
\cite{bennett64} and Diamond et al. \cite{diamond64}. 

In the field of theoretical neuroscience, Hebb \cite{hebb49} wrote his ideas in
words that inspired mathematical modelling related to synaptic plasticity
\cite{gerstner12}. According to Hebbian theory, the synaptic strength increases
when a presynaptic neuron participates in the firing of a postsynaptic neuron, 
in other words, neurons that fire together, also wire together. The Hebbian
plasticity led to the modelling of spike timing-dependent plasticity (STDP)
\cite{markram12,borges16}. It was possible to obtain the STDP function for
excitatory synapses by means of synaptic plasticity experiments performed by Bi
and Poo \cite{poo98}. The STDP function for inhibitory synapses was reported in
experimental results in the entorthinal cortex by Haas et al. \cite{haas06}.

In this review, we show results that allow to understand the relation
between spike synchronisation and synaptic plasticity and this dependence with
the non-trivial topology that is induced in the brain due to STDP. As so, we
consider an initial all-to-all network, where the neuronal network is built by
connecting neurons by means of excitatory and inhibitory synapses. We show
that the transition from weakly synchronous to strongly synchronous states 
depends on the neuronal network architecture, as well as to the STDP network 
evolves to non-trivial topology. When the strength of the inhibitory 
connections is of the same order of that of the excitatory connections, the 
final topology in the plastic brain presents the rich-club phenomenon, where 
neurons that have high degree connectivity towards neurons of the same 
presynaptical group (either excitatory of inhibitory) become strongly connected
to neurons of the other postsynaptical group. The final topology has all the 
features of a non-trivial topology, when the strength of the synapses becomes 
reasonably larger than the strength of the excitatory connections, where 
neurons only sparsely connect to other neurons.  

The structure of the review is the following. In Section $2$, we introduce the
Hodgkin-Huxley model for a neuron and the synchronisation dynamics of neuronal
networks. Section $3$ presents the Hebbian rule and the spike-timing dependent
plasticity (STDP) in excitatory and inhibitory syna\-pses. In Section $4$, we
show the effects of the synaptic plasticity on the network topology and
synchronous behaviour. Finally, in the last Section, we draw the conclusions.


\section{Hodgkin-Huxley Neuronal Networks}

\subsection{Neurons}

\begin{figure}[htbp]
\begin{center}
\includegraphics[height=7cm,width=9cm]{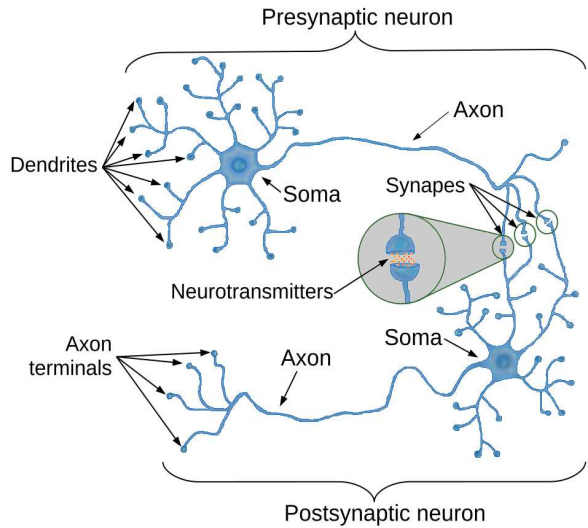}
\caption{Schematic illustration showing the three main parts of neurons
(dendrite, soma and axon), including the presynaptic and postsynaptic neurons.}
\label{fig1}
\end{center}
\end{figure}

Neurons are cells responsible for receiving, processing and transmitting
information in the neuronal system \cite{alberts02}. They have differences
in sizes, length of axons and dendrites, in the number of dendrites and axons
terminals. Figure \ref{fig1} illustrates the three main parts of the neuron:
dendrite, cell body or soma, and axon \cite{arbib02}. The dendrites are
responsible for the signal reception, and the axons drive the impulse from the
cell body to another neuron. The neurons are connected through synapses, where
the neuron that sends the signal is called presynaptic and the postsynaptic is
the neuron that receives it. The most common form of neuron communication is by
means of the chemical synapses, where the signal is propagated from the
presynaptic to postsynaptic neurons by releasing neurotransmitters.

\begin{figure}[htbp]
\begin{center}
\includegraphics[height=5cm,width=8cm]{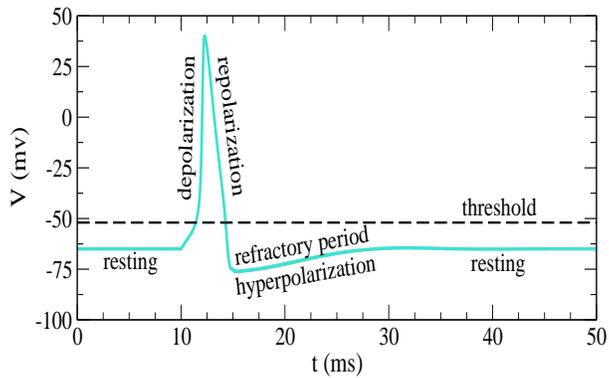}
\caption{Plot of the action potential showing the various phases at a point on
the cell membrane.}
\label{fig2}
\end{center}
\end{figure}

The signal propagates by means of the variation of internal neuron electric
potential. An action potential occurs when a neuron sends information from the
soma to the axon. The action potential is characterised by a rapid
change in the membrane potential, as shown in Fig. \ref{fig2}. In the absence
of stimulus, the membrane potential remains near a baseline level. A
depolarisation occurs when the action potential is greater than a threshold
value. After the depolarisation, the action potential goes through a certain
repolarisation stage, where the action potential rapidly reaches the
refractory period or hyperpolarisation. The refractory period is the time
interval in which the axon does not transmit the impulse \cite{arbib02}.

\begin{figure}[htbp]
\begin{center}
\includegraphics[height=6cm,width=8cm]{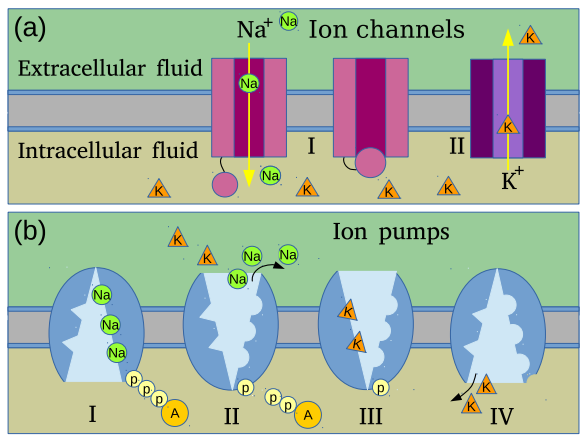}
\caption{Schematic diagram of the ions traffic across cell membranes, (a) ion
channels and (b) ion pumps.}
\label{fig3}
\end{center}
\end{figure}

Action potentials are generated and propagates due to different ions crossing
the neuron membrane. The ions can cross the membrane through ion channels and
ion pumps \cite{gouaux05}. Figure \ref{fig3}(a) shows the ion channels of
sodium (${\rm Na}^+$) and potassium (${\rm K}^+$). In the depolarisation stage,
a great amount of sodium ions move into the axon (I), while the repolarisation
occurs when the potassium ions move out of the axon (II). Figure \ref{fig3}(b)
shows the transport of sodium (I and II) and potassium ions (III and IV)
through the pumps. The sodium-potassium pumps transport sodium ions out and
potassium ions in, and it is responsible for maintaining the resting potential
\cite{gouaux05}.

\subsection{Hodgkin-Huxley Model}

\begin{figure}[htbp]
\begin{center}
\includegraphics[height=5cm,width=8cm]{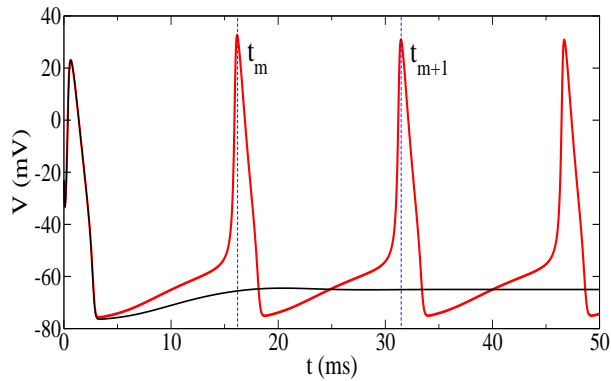}
\caption{\small Membrane potential $V$ of a Hodgkin-Huxley neuron with
$I=0\mu$A/cm$^2$ (black line) and $I=9\mu$A/cm$^2$ (red line).}
\label{fig4}
\end{center}
\end{figure}

Hodgkin and Huxley \cite{hodgkin52} performed experiments on the giant squid
axon using microelectrodes introduced into the intracellular medium. They
proposed a mathematical model that allowed the development of a quantitative
approximation to understand the biophysical mechanism of action potential
generation. In 1963, Hodgkin and Huxley were awarded with the Nobel Prize in
Physiology or Medicine for their work. The Hodgkin-Huxley model is given by
\begin{eqnarray}\label{Equacoes_HH}
C\dot{V} & = & I-g_{\rm K}n^{4}(V-E_{\rm K})-g_{\rm Na}m^{3}h(V-E_{\rm Na}) -g_{L}(V-E_{\rm L}), \nonumber \\
\dot{n} & = & \alpha_{n}(V)(1-n)-\beta_{n}(V)n,\\
\dot{m} & = & \alpha_{m}(V)(1-m)-\beta_{m}(V)m,\\
\dot{h} & = & \alpha_{h}(V)(1-h)-\beta_{h}(V)h,
\end{eqnarray}
where $C$ is the membrane capacitance ($\mu$F/cm$^2$), $V$ is the membrane 
potential (mV), $I$ is the constant current density, parameter $g$ is the
conductance, and $E$ the reversal potentials for each ion. The functions $m(V)$
and $n(V)$ represent the activation for sodium and potassium, respectively, and
$h(V)$ is the function for the inactivation of sodium. The functions
$\alpha_{n}$, $\beta_{n}$, $\alpha_{m}$, $\beta_{m}$, $\alpha_{h}$, $\beta_{n}$
are given by
\begin{eqnarray}
\alpha_{n}(v) & = & \frac{0.01 v + 0.55}{1 - \exp \left(-0.1 v-5.5 \right)},\\
\beta_{n}(v) & = & 0.125\exp\left(\frac{-v-65}{80}\right),\\
\alpha_{m}(v) & = & \frac{0.1 v + 4}{1 - \exp\left (-0.1 v - 4\right)},\\
\beta_{m}(v) & = & 4\exp\left(\frac{-v-65}{18}\right),\\
\alpha_{h}(v) & = & 0.07\exp\left(\frac{-v-65}{20}\right),\\
\beta_{h}(v) & = & \frac{1}{1 + \exp\left(-0.1 v - 3.5\right)},
\end{eqnarray}
where $v=V/$[mV]. We consider $C=1$ $\mu$F/cm$^{2}$, $g_{\rm K}=36$mS/cm$^{2}$,
$E_{\rm K}=-77$mV, $g_{\rm Na}=120$mS/cm$^{2}$, $E_{\rm Na}=50$mV, 
$g_{\rm L}=0.3$mS/cm$^{2}$, $E_{\rm L}=-54.4$mV \cite{borges17}. Depending
on the value of the external current density $I$ ($\mu$A/cm$^2$) the neuron can 
present periodic spikings or single spike activity. In the case of periodic 
spikes, if the constant $I$ increases, the spiking frequency also increases.
Figure \ref{fig4} shows the temporal evolution of the membrane potential of a 
Hodgkin-Huxley neuron for $I=0\mu$A/cm$^2$ (black line) and for
$I=9\mu$A/cm$^2$ (red line). For the case without current, the neuron shows
an initial firing and, after the spike, it remains in the resting potential. In
the second case the external current $I$ is greater than the required threshold
and the neuron exhibits firings.

\subsection{Neuronal Synchronisation}

The synchronisation process here is related to natural phenomena ranging from
metabolic processes in our cells to the highest cognitive activities
\cite{arenas08}. Neuronal synchronisation has been found in the brain during
different tasks and at rest \cite{deco11}. We study in this text neuronal
synchronisation process in a network of coupled Hodgkin-Huxley neurons. The
network dynamics is given by
\cite{popovych13}
\begin{eqnarray}
C\dot{V_{\rm i}} & = & I_{\rm i}-g_{\rm K}n^{4}(V_{\rm i}-E_{\rm K})-
g_{\rm Na}m^{3}h(V_{\rm i}-E_{\rm Na}) \nonumber \\
& & - g_{L}(V_{\rm i}-E_{\rm L})+\frac{(V_{\rm r}^{\rm Exc}-
V_{\rm i})}{\omega_{\rm Exc}} \sum_{j=1}^{N_{\rm Exc}}\varepsilon_{\rm ij}s_{\rm j}
\nonumber \\
& & +\frac{(V_{r}^{\rm Inhib}-V_{\rm i})}{\omega_{\rm Inhib}}\sum_{j=1}^{N_{\rm Inhib}}
\sigma_{\rm ij}s_{\rm j}+\Gamma_i,
\end{eqnarray}
where the elements of the matrix $\varepsilon_{\rm ij}$ ($\sigma_{\rm ij}$) are
the intensity of the  excitatory (inhibitory) synapse (coupling strength)
between the presynaptic neuron $j$ and the postsynaptic neuron $i$,
$\omega_{\rm Exc}$ ($\omega_{\rm Inhib}$) represents the me\-an number of excitatory
(inhibitory) synapses of each neuron, $\Gamma_i$ is an external perturbation so
that the neuron is randomly chosen and the chosen one receives an input with a
constant intensity $\gamma$, $N_{\rm Exc}$ is the number of excitatory neurons,
and $N_{\rm Inhib}$ is the number of inhibitory neurons. The excitatory
(inhibitory)neurons are connected with reverse potential
$V_{\rm r}^{\rm Exc}=20 \rm mV$ ($V_{\rm r}^{\rm Inhib}=-75\rm mV$), and the
postsynaptic potential $s_{i}$ is given by \cite{popovych13}
\begin{equation}
\frac{ds_{i}}{dt}=\frac{5(1-s_{i})}{1+\exp(-\frac{V_{i}+3}{8})}-s_{i}.
\end{equation}

One measure that we adopt to quantify synchronous behaviour is the Kuramoto 
order parameter that reads as \cite{kuramoto84}
\begin{equation}
Z(t) = R(t){\rm e}^{i\psi(t)}= \frac{1}{N}\sum_{j=1}^{N}{\rm e}^{i\theta_{j}(t)},
\end{equation}   
where $R(t)$ is the amplitude, $\psi(t)$ is the angle of a centroid phase
vector, and
\begin{equation}
\theta_{j}(t)=2\pi\frac{t-t_{j,m}}{t_{j,m+1}-t_{j,m}}
\end{equation}
is the phase of the neuron $j$, with $t_{j,m}< t < t_{j,m+1}$. The time $t_{j,m}$
denotes the $m$-th spike of the neuron $j$. In a complete synchronised state the
network exhibits $R=1$. For a strongly synchronised regime it has
$R\geq 0.9$, whereas a weakly synchronous behaviour occurs for $R<0.9$.

\begin{figure}[htbp]
\begin{center}
\includegraphics[height=7cm,width=8cm]{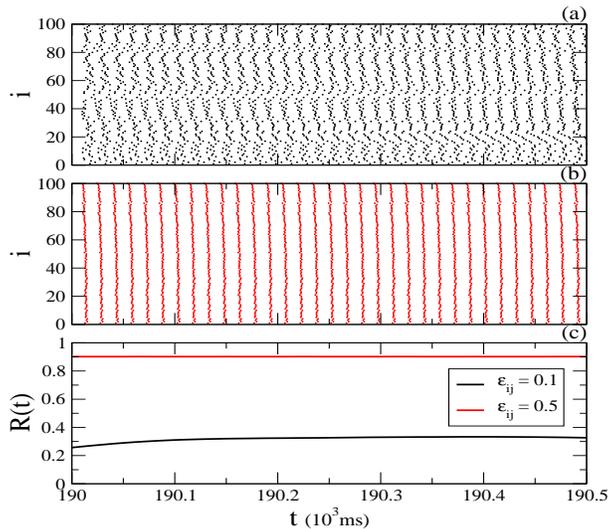}
\caption{(Colour online) Raster plots of spike onsets for a random network
with $100$ Hodgkin-Huxley neurons, $\gamma=0$, (a) $\varepsilon_{\rm ij}=0.1$ and
(b) $\varepsilon_{\rm ij}=0.5$. In (c) the time evolution of the Kuramoto order
parameter for $\varepsilon_{\rm ij}=0.1$ (black line) and
$\varepsilon_{\rm ij}=0.5$ (red line).}
\label{fig5}
\end{center}
\end{figure}

Figure \ref{fig5}(a) and (b) exhibit the raster plots of spike onsets for a
random network with $100$ Hodgkin-Huxley neurons coupled by means of excitatory
synapses, mean degree $K=10$, $\gamma=0$, excitatory coupling intensity
$\varepsilon_{\rm ij}=0.1$ and $\varepsilon_{\rm ij}=0.5$, respectively. In Figure
\ref{fig5}(a), the neuronal network presents weakly synchronous behaviour, 
while in Figure \ref{fig5}(b) the network shows stron\-gly synchronised spiking 
(though not complete synchronisation). Figure \ref{fig5}(c) shows the order 
parameter $R(t)$ for $\varepsilon_{\rm ij}=0.1$ (black line) and 
$\varepsilon_{\rm ij}=0.5$ (red line). By increasing the coupling strength, from
$0.1$ to $0.5$, the neuronal network asymptotes to a synchronous behaviour.

\section{Spike-Timing Dependent Plasticity}

Work carried to try to unveil the role of synaptic plasticity in learning and
memory has the Hebb rule as a basis. Hebb rule is a postulate proposed in 1949
by Hebb in his book ``The organization of behavior'' \cite{hebb49}. He
conjectured that the synapse from presynaptic to postsynaptic neuron
should be maximally strengthened if the input from presynaptic neuron
contributes to the firing of postsynaptic. In this way, a long-term
potentiation is caused when there is coincident spiking of presynaptic and
postsynaptic neurons \cite{gerstner10}.

In the synaptic plasticity, synapse weakening and strengthening are implemented
by long-term depression (LTD) and potentiation (LTP), respectively
\cite{feldman12}. LTP refers to a long-lasting increase in excitatory
postsynaptic potential, while LTD decreases the efficacy of a synapse. Bliss et
al. \cite{bliss73} suggested that low-frequency firing drives LTD, whereas LTP
is driven by presynaptic firing of the high-frequency. Synaptic plasticity 
alteration as a function of the relative timing of presynaptic and postsynaptic
firing was named as spike timing-dependent plasticity (STDP) by Song et al. 
\cite{song00}. STDP has been observed in brain regions, and relevant studies on
it were carried out by Gerstner \cite{gerstner96} and Markram et al.
\cite{markram95,markram97}. Fr\'egnac et al. \cite{fregnac10} provided the
existence of STDP in cat visual cortex {\it in vivo}. Moreover, research on
STDP has focused in the hippocampus and cortex \cite{buchanan10}.

We have studied the changes in synchronous and desynchronous states caused in a
Hodgkin-Huxley network due to excitatory (eSTDP), as well as inhibitory (iSTDP)
spike timing-dependent plasticity. We have considered the plasticity as a
function of the difference of postsynaptic and presynaptic excitatory and
inhibitory firing according to Refs. \cite{poo98} and \cite{haas06},
respectively.

The excitatory eSTDP is given by
\begin{equation}\label{eqplast}
\Delta \varepsilon_{ij}= \left\{
\begin{array}{ll}
\displaystyle A_{1}\exp(-\Delta t_{ij}/\tau_{1})\;,\;\Delta t_{ij}\geq 0 \\
\displaystyle -A_{2}\exp(\Delta t_{ij}/\tau_{2})\;,\;\Delta t_{ij} < 0 
\end{array}
\right. ,
\end{equation}
where
\begin{equation}
\Delta t_{ij}=t_{i}-t_{j}=t_{\rm pos}-t_{\rm pre},
\end{equation}
$t_{\rm pos}$ is the spike time of the postsynaptic neuron, and $t_{\rm pre}$ is
the spike time of the presynaptic one.

\begin{figure}[htbp]
\begin{center}
\includegraphics[height=7cm,width=8cm]{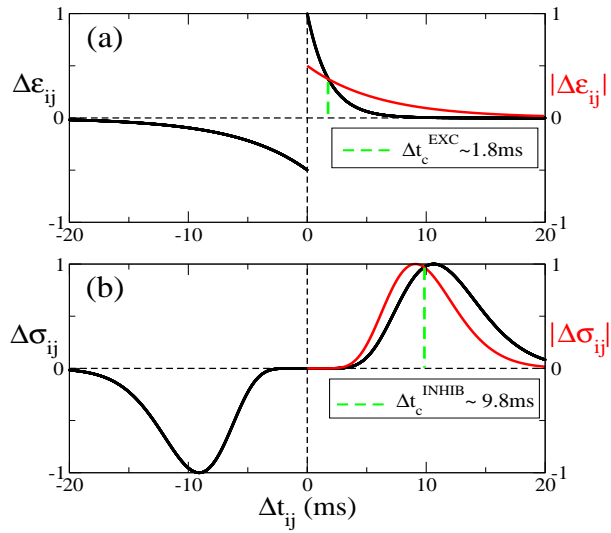}
\caption{Plasticity as a function of the difference of spike timing of
postsynaptic and presynaptic synapses for (a) excitatory (eSTDP) and (b)
inhibitory (iSTDP). The green dashed line indicates the intersection between
the potential and depression curves.}
\label{fig6}
\end{center}
\end{figure}

Figure \ref{fig6}(a) shows the result obtained from Eq. (\ref{eqplast}) for
$A_{1}=1.0$, $A_{2}=0.5$, $\tau_{1}=1.8$ms, and $\tau_{2}=6.0$ms. The initial
synaptic weights $\varepsilon_{ij}$ are normally distributed with mean and
standard deviation equal to $0.25$ and $0.02$, respectively
($0\leq \varepsilon_{ij}\leq 0.5$). They are updated according to Eq.
(\ref{eqplast}), where
\begin{equation}
\varepsilon_{ij}\rightarrow \varepsilon_{ij}+10^{-3}\Delta\varepsilon_{ij}.
\end{equation}  
The green dashed line denotes the intersection between the absolute values of
the depression (black line) and potentiation (red line) curves. For
$\Delta t_c^{Exc}<1.8$ms the potentiation is larger than the depression. In 
addition, the red line denotes the absolute value of the coupling strength
($|\Delta \varepsilon_{ij}|$).

In the inhibitory iSTDP synapses, the coupling str\-ength $\sigma_{ij}$ is 
adjusted according to the equation
\begin{equation}\label{eqplastI}
\Delta \sigma_{ij} =  \frac{g_0}{g_{\rm norm}} {\alpha}^{\beta} |\Delta t_{ij}| 
{\Delta t_{ij}}^{\beta -1} \exp(-\alpha |\Delta t_{ij}|),
\end{equation}
where $g_0$ is the scaling factor accounting for the amount of change in 
inhibitory conductance induced by the synaptic plasticity rule, and
$g_{\rm norm} = {\beta}^{\beta}  \exp(-\beta)$ is the normalising constant.
In Figure \ref{fig6}(b) we see the result obtained from Eq. (\ref{eqplastI})
for $g_0 = 0.02$, $\beta=10.0$, $\alpha =0.94$ if $\Delta t_{ij}>0$, and for
$\alpha=1.1$ if $\Delta t_{ij}<0$. As a result, $\Delta \sigma_{ij}>0$ for
$\Delta t_{ij}>0$, and $\Delta \sigma_{ij}<0$ for $\Delta t_{ij}<0$. The initial
inhibitory synaptic weights $\sigma_{ij}$ are normally distributed with mean and
standard deviation equal to $\sigma=c\varepsilon$ ($1\leq c\leq 3$) and 0.02,
respectively ($0\leq \sigma_{ij}\leq 2c\varepsilon$). The coupling strengths are
updated according to Eq. (\ref{eqplastI}), where
\begin{equation}
\sigma_{ij}\rightarrow \sigma_{ij}+10^{-3}\Delta\sigma_{ij}.
\end{equation}
The updates for $\varepsilon_{ij}$ and $\sigma_{ij}$ are applied to the last
postsynaptic spike. For $\Delta t_c^{\rm Inhib}<9.8$ms the depression is larger
than the potentiation.


\section{Influence of the Synaptic Plasticity on the Network Topology}

\subsection{Without External Perturbation}

About $20\%$ of the synapses in the brain have inhibitory characteristics
\cite{noback05}. We consider that the intensities of both excitatory and
inhibitory synapses are modifiable over time by a plasticity rule. We use
a network of $200$ Hodgkin-Huxley neurons with $I_i$ normally distributed in the
interval [$9.0$-$10.0$]. $E_i$ represents the $i$-th excitatory neurons with 
sub-index $i$ in the interval [$1$-$160$] and $I_i$ represents the $i$-th 
inhibitory neuron with the sub-index $i$ in [$161$-$200$]. In all the 
simulations, we consider a total time interval of $2000$s.

\begin{figure}[htbp]
\begin{center}
\includegraphics[height=7cm,width=8cm]{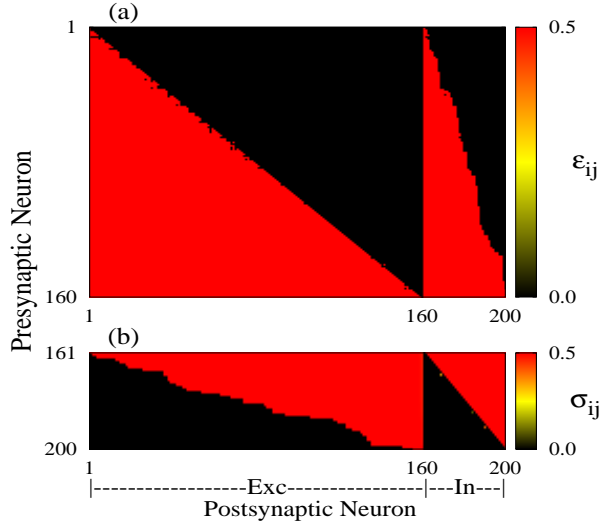}
\caption{Intensity of the final coupling for initial couplings with
$\frac{\sigma}{\varepsilon}=1$ and $\gamma=0$, (a) excitatory and (b) 
inhibitory synapses. The coupling matrix has a triangular shape.}
\label{fig7}
\end{center}
\end{figure}

When the initial intensity of the inhibitory synapses is small
($\frac{\sigma}{\varepsilon} \approx 1$), we show that the potentiation
occurs in both kinds of synapses and the final coupling matrix exhibits a
triangular shape, as seen in Fig. \ref{fig7}. In the excitatory synapses a
reinforcement is observed from the neurons of greater to smaller frequency
(Fig. \ref{fig7}(a)), whereas in the inhibitory synapses, the potentiation
occurs from the neurons of smaller to greater frequency (Fig. \ref{fig7}(b)). 
Figure \ref{fig7}(a) points out that presynaptic excitatory neurons that are 
more likely to strongly connect to a large number of postsynaptic excitatory 
neurons are also more likely to strongly connect to postsynaptic inhibitory 
neurons. Similarly,  Figure \ref{fig7}(b) points out that presynaptic 
inhibitory neurons that are more likely to strongly connect to a large number 
of postsynaptic inhibitory neurons are also more likely to stron\-gly connect 
to postsynaptic excitatory  neurons. This reveal a rich club phenomenon in the 
neural plasticity, where the neurons with larger degrees to its own "club" 
(either the excitatory or the inhibitory community) tend to be also more 
connected to the other "club".  The rich-club phenomenon is know to exist in 
the topological organisation of the brain \cite{towlson13} and was recently 
hypothetised to be an effect of Hebbian learning mechanisms in Ref.
\cite{vertes14}.

\begin{figure}[htbp]
\begin{center}
\includegraphics[height=6cm,width=7cm]{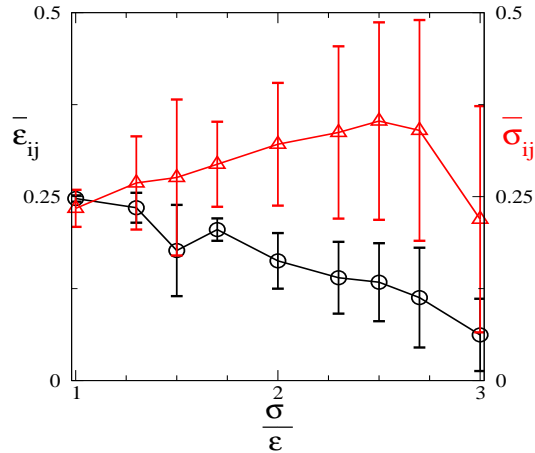}
\caption{Mean excitatory (black circles) and inhibitory (red triangles)
couplings as a function of $\frac{\sigma}{\varepsilon}$, where we consider
simulations without external perturbations. The bars indicate the standard
deviation calculated for the mean value from 30 simulations.}
\label{fig8}
\end{center}
\end{figure}

\begin{figure}[htbp]
\begin{center}
\includegraphics[height=7cm,width=8cm]{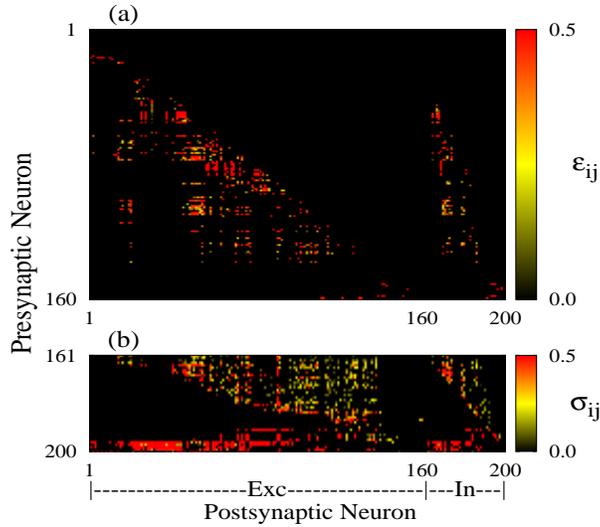}
\caption{Intensity of the couplings for $\frac{\sigma}{\varepsilon}=2.7$,
$\gamma=0$, $t=2000$s, (a) excitatory and (b) inhibitory synapses. The network
has disconnected neurons.}
\label{fig9}
\end{center}
\end{figure}

\begin{figure}[htbp]
\begin{center}
\includegraphics[height=7cm,width=8cm]{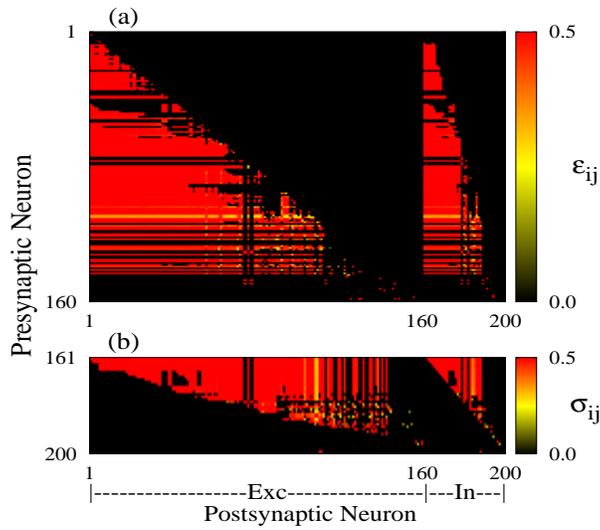}
\caption{Intensity of the couplings for $\frac{\sigma}{\varepsilon}=2.7$,
$\gamma=0$, $t=2000$s, (a) excitatory and (b) inhibitory synapses. The neural
network contains disconnected blocks.}
\label{fig10}
\end{center}
\end{figure}

In Fig. \ref{fig8} it is exhibited the value of the excitatory
$(\bar{\varepsilon})$ and the inhibitory $(\bar{\sigma})$ mean coupling as a
function of $\frac{\sigma}{\varepsilon}$. A small variability around the mean
values of the excitatory and inhibitory couplings is observed for small values
of $\frac{\sigma}{\varepsilon}$. However, increasing the inhibitory synapse
implies in an increase in the variability around both mean values, as
indicated by the standard deviation bars. This fact becomes notable when the
initial intensity of the inhibitory synapses is greater than
$\frac{\sigma}{\varepsilon}=1.5$. As a result, the inhibitory synapses act more
intensely on the neuronal network dynamics, and a different asymptotic
behaviour can be observed. Figures \ref{fig9} and \ref{fig10}, at $t=2000$s,
show the coupling matrices with the values of the excitatory and inhibitory
couplings for an initial value given by $\frac{\sigma}{\varepsilon}=2.7$. In 
some simulations, the synaptic connections tend to zero, namely, the network 
becomes disconnected (Fig. \ref{fig9}). In other simulations, disconnected 
blocks are observed, as shown in Fig. \ref{fig10}. Nevertheless, for the same 
value of the $\frac{\sigma}{\varepsilon}$ parameter, the system can exhibit an 
asymptotic behaviour similar to the case when initial coupling have
$\frac{\sigma}{\varepsilon}=1.0$ (Fig.\ref{fig7}).

The behaviour observed in the synapse intensity can be explained in terms of
the average time between spikes. For that, we defined the mean time between
spikes among neurons having both excitatory and inhibitory synapses by the
equations  
\begin{eqnarray}
\bar{\Delta t}_{ij}^{\rm Exc} & = & \frac{1}{\tau}\sum_{i\neq j}
| t_{\rm pre}^{\rm Exc}-t_{\rm pos} |, \\
\bar{\Delta t}_{ij}^{\rm Inhib} & = & \frac{1}{\tau}\sum_{i\neq j}
| t_{\rm pre}^{\rm Inhib}-t_{\rm pos} |. 
\end{eqnarray}

\begin{figure}[htbp]
\begin{center}
\includegraphics[height=7cm,width=7cm]{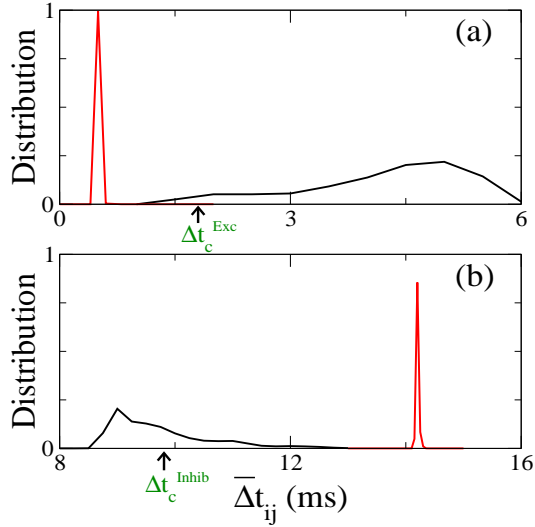}
\caption{Probability distribution frequency of the average firing times for
$\frac{\sigma}{\varepsilon}=2.7$, $\gamma=0$, (a) excitatory and (b) inhibitory
synapses. For the triangular shape and unidirectionally connected coupling
matrix (Fig. \ref{fig7}), the $\bar{\Delta t_{ij}}$ values are more frequently
found in the potentiation regions (red curves in (a) and (b)). The black lines
in (a) and (b) illustrate the completely opposite case observed in Fig.
\ref{fig9}. The values of $\Delta t_c$ were obtained in Fig. \ref{fig6}.}
\label{fig11}
\end{center}
\end{figure}

In Figure \ref{fig11}, $\bar{\Delta t}_{ij}^{\rm Exc}$ and 
$\bar{\Delta t}_{ij}^{\rm Inhib}$ values are show for the extreme case of initial
couplings given by $\frac{\sigma}{\varepsilon}=2.7$ (black lines) and initial 
coupling given by $\frac{\sigma}{\epsilon}=1.0$ (red lines). For the case where
the neuronal network becomes disconnected (black lines), the average time 
values that are more frequently are found in the depression region of the eSTDP
and iSTDP models ($\bar{\Delta t}_{ij}^{\rm Exc}>\Delta t_{c}^{\rm Exc}=1.8\rm ms$ 
and $\bar{\Delta t}_{ij}^{\rm Inhib}<\Delta t_{c}^{\rm Inhib}=9.8\rm ms$). However, in
simulations where a neuronal network becomes strongly connected, a higher 
concentration of the average time values in the potentiation regions of the 
plasticity models is observed
($\bar{\Delta t}_{ij}^{\rm Exc}<\Delta t_{c}^{\rm Exc}=1.8\rm ms$ and
$\bar{\Delta t}_{ij}^{\rm Inhib}>\Delta t_{c}^{\rm Inhib}=9.8\rm ms$). So, 
potentiation happening for high frequencies excitatory synapses and lower 
frequencies inhibitory synapses promote the strengthening of synaptic 
connectivity and the rich-club phenomenon.

\subsection{With External Perturbation}

\begin{figure}[htbp]
\begin{center}
\includegraphics[height=7cm,width=8cm]{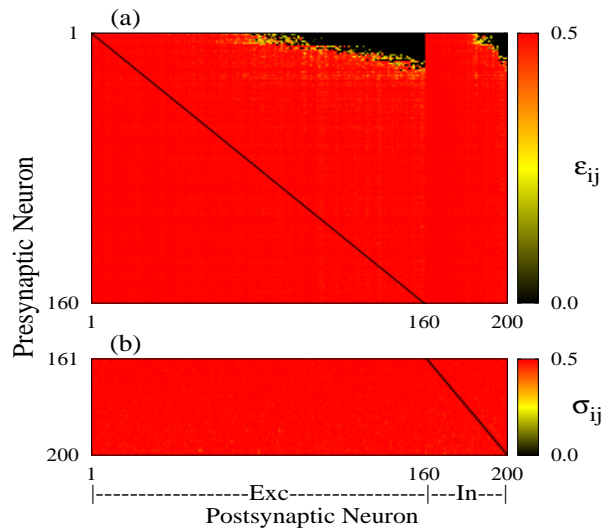}
\caption{Perturbed intensity of the final coupling for
$\frac{\sigma}{\varepsilon}=2.2$, $\gamma=10\mu\rm A/cm^2$, (a) excitatory and
(b) inhibitory synapses. Almost all connections in the neuronal networks are
reinforced.}
\label{fig12}
\end{center}
\end{figure}

An external perturbation combined with eSTDP and iSTDP can provide a positive
contribution to the excitatory and inhibitory mean coupling. In this case, we
observe that when the influence of the inhibitory is smaller than the
excitatory synapse ($\frac{\sigma}{\varepsilon}<2.3$), the potentiation occurs
in approximately all the synapses (excitatory and inhibitory) (Fig.
\ref{fig12}). Then, the network remains stron\-gly connected, with a topology
close to all-to-all. Almost all the intensities of the connections converge to
high values (${\bar\varepsilon}_{ij}\geq 0.4$ and
${\bar\sigma}_{ij}\approx 0.5$). Only a few connections, where the presynaptic
neurons have lower frequency, tend to zero.

\begin{figure}[htbp]
\begin{center}
\includegraphics[height=7cm,width=8cm]{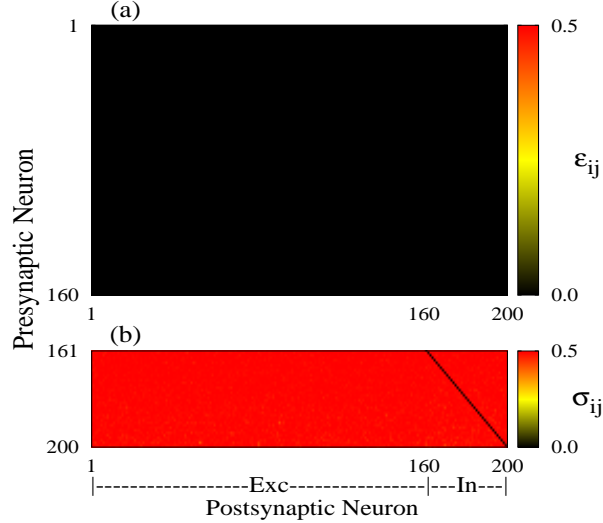}
\caption{Perturbed intensity of the final coupling for for initial coupling 
given by $\frac{\sigma}{\varepsilon}=2.4$, $\gamma=10\mu\rm A/cm^2$, (a) 
excitatory and (b) inhibitory synapses. All the excitatory connections in the 
neuronal networks disappear, but the inhibitory synapses are enhanced.}
\label{fig13}
\end{center}
\end{figure}

For larger $\frac{\sigma}{\varepsilon}$ values, we also observe that the
inhibi\-tory connections become strengthened. The inhibitory mean coupling
converges to the largest value allowed in the interval when
$\frac{\sigma}{\varepsilon}>2.3$. However, for this same value of
$\frac{\sigma}{\varepsilon}$, there is a trend of decreasing intensity of
excitatory synapses ($\bar{\varepsilon_{ij}}\approx 0$). The neurons remain
connected through the inhibitory synapses (Fig. \ref{fig13}). 

\begin{figure}[htbp]
\begin{center}
\includegraphics[height=6cm,width=7cm]{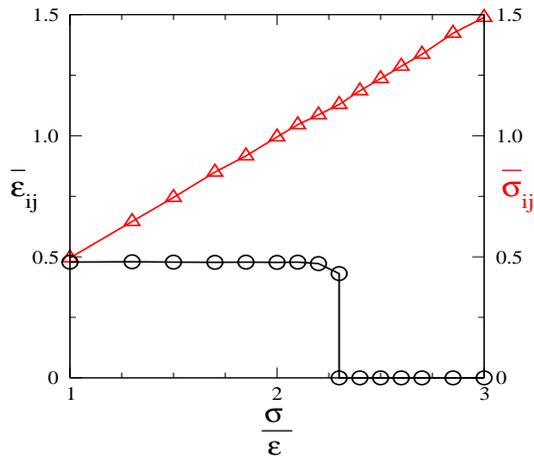}
\caption{Perturbed mean excitatory (black circles) and inhibitory (red
triangles) couplings as a function of $\frac{\sigma}{\varepsilon}$, where we
consider $\gamma=10\mu\rm A/cm^2$.}
\label{fig14}
\end{center}
\end{figure}

An abrupt transition in the mean excitatory coupling values can also be
seen for $\frac{\sigma}{\varepsilon} \approx 2.3$. For values slightly less
than $2.3$ ($\frac{\sigma}{\varepsilon}=2.2$), both excitatory and inhibitory
synapses undergo an increase in their intensities, whereas, for values
of $\frac{\sigma}{\varepsilon}$ larger than this threshold, the inhibitory
synapses undergo potentiation while the excitatory sy\-napses tend to zero
(Fig. \ref{fig14}).

\begin{figure}[htbp]
\begin{center}
\includegraphics[height=7cm,width=7cm]{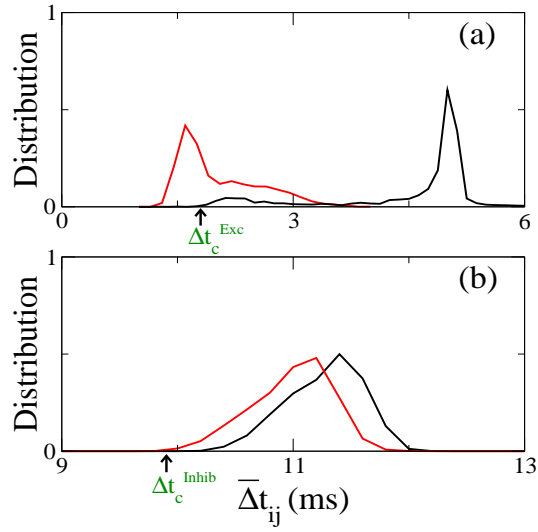}
\caption{Probability distribution function of the average firing times for
$\frac{\sigma}{\varepsilon}=2.3$, $\gamma=10\mu\rm A/cm^2$, (a) excitatory and
(b) inhibitory synapses. For all-to-all topology (Fig \ref{fig12},
the $\bar{\Delta t_{ij}}$ vales are more frequently found in the potentiation
regions (red curves in (a) and (b)). The black lines in (a) and (b)
illustrate the completely opposite case observed in Fig. \ref{fig13}. The
values of $\Delta t_c$ were obtained from Fig. \ref{fig6}.}
\label{fig15}
\end{center}
\end{figure}

The time evolution of both excitatory and inhibitory synapses depend on the
time interval between spikes of presynaptic and postsynaptic neurons. Figure
\ref{fig15} shows the frequency between the mean times among presynaptic and
postsynaptic spikes. This figure exhibits the two extreme cases, when the
neuronal network converges to a stron\-gly connected global topology or to a
network with only inhibitory synapses, for $\frac{\sigma}{\varepsilon}=2.3$.
When the increase of the weights occurs in almost all the synapses, the
$\bar{\Delta t_{ij}}$ values  appear more frequently in the regions of
potentiation of both models of plasticity
($\bar{\Delta t}_{ij}^{\rm Exc}<\Delta t_{c}^{\rm Exc}=1.8\rm ms$ and
$\bar{\Delta t}_{ij}^{\rm Inhib}>\Delta t_{c}^{\rm Inhib}=9.8\rm ms$). However, when
only strong inhibitory synapses are observed in the final neuronal network, it
is verified that $\bar{\Delta t_{ij}}$ values in excitatory synapses are more
frequently found in the depression region of the eSTDP model
($\bar{\Delta t}_{ij}^{\rm Exc}>\Delta t_{c}^{\rm Exc}=1.8\rm ms$). In this case,
the inhibitory synapses are reinforced due to the fact that the
$\bar{\Delta t_{ij}}$ values are more frequently found in the region of
potentiation of the iSTDP model
(${\Delta t}_{ij}^{\rm Inhib}>\Delta t_{c}^{\rm Inhib}=9.8\rm ms$).

Therefore, noise can always enhance inhibitory sy\-napses in the plastic brain. 
Excitatory synapses can also be enhanced if the initial network has 
sufficiently large excitatory synaptic strength (no less than about half the 
value of the inhibitory synapses strength).

\section{Influence of the Synaptic Plasticity on the Synchronous Behaviour}

\subsection{Without External Perturbation}

\begin{figure}[htbp]
\begin{center}
\includegraphics[height=6cm,width=7cm]{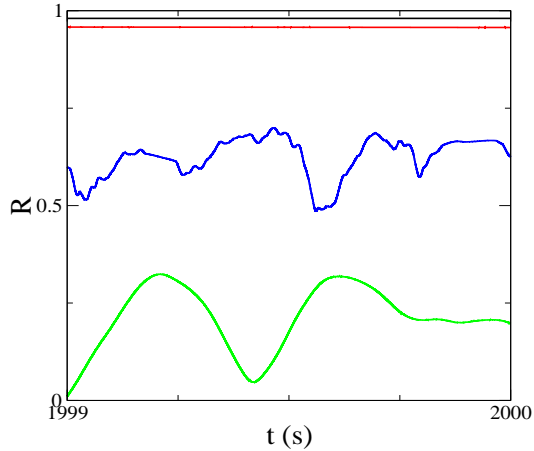}
\caption{Order parameter for $\frac{\sigma}{\varepsilon}=1.0$ (black line) and
$\frac{\sigma}{\varepsilon}=2.7$ (red, blue and green lines).}
\label{fig16}
\end{center}
\end{figure}

The change in the behaviour of the synapse intensity between presynaptic and
postsynaptic neurons due to plasticity is reflected on the spike
synchronisation. In Fig. \ref{fig16} we observe different behaviours in
relation to synchronisation, where we calculate the order parameter. Figure
\ref{fig16} exhibits the behaviour of the order parameter as a function of time
for simulations without external perturbations, discarding a large transient
ti\-me. The neuronal network evolves to the strong synchronised state with
$R(t)>0.9$ (black line) if the initial ratio of intensities of the inhibitory
synapses are weak ($\frac{\sigma}{\varepsilon}\approx 1.0$), this inhibition and
excitation have similar initial strengths. However, with the increase
of the inhibitory synapses intensities $\frac{\sigma}{\varepsilon}>1.5$, 
different final states are observed in relation to the synchronisation (red, 
green and blue lines). 

\subsection{With External Perturbation}

\begin{figure}[htbp]
\begin{center}
\includegraphics[height=6cm,width=7cm]{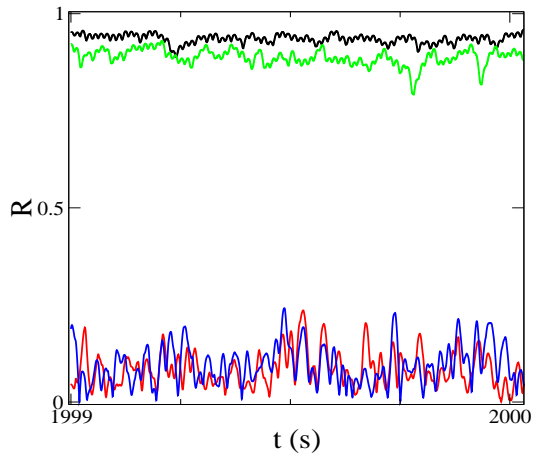}
\caption{Order parameter for $\gamma=10\mu\rm A/cm^2$,
$\frac{\sigma}{\varepsilon}=1.0$ (black line), $\frac{\sigma}{\varepsilon}=3.0$
(red line) and $\frac{\sigma}{\varepsilon}=2.3$ (blue and green lines).}
\label{fig17}
\end{center}
\end{figure}

We consider an external perturbation ($\gamma=10\mu\rm A/cm^2$) when the
initial inhibitory synapses intensity ratio are small
($\frac{\sigma}{\varepsilon}\approx 1.0$). In this case, the network has a
synchronous behaviour ($\bar{R}(t)>0.9$), as shown in Fig. \ref{fig17} (black
line). When inhibitory synapses intensities have a great influence on the
network dynamics ($\frac{\sigma}{\varepsilon}\approx 3.0$), neurons tend to
exhibit desynchronised firing behaviour with $\bar{R}(t) \approx 0.1$ (red
line). However, when $\frac{\sigma}{\varepsilon}\approx 2.3$, we observe two
possible asymptotic values for the order parameter. In some simulations a 
strongly synchronised behaviour appears, while in others it is observed a 
weakly synchronous evolution of spikes between the neurons in the network 
(green and blue lines).

\section{Conclusions}

Neuronal networks based on the Hodgkin-Huxley mo\-del have been used to simulate
coupled spiking neurons. The Hodgkin-Huxley neuron is a coupled set of ordinary
nonlinear differential equations that describes the ionic basis of the membrane
potential. In this review, we considered a Hodgkin-Huxley network with synaptic
plasticity (STDP). The STDP is a process that adjusts the strength of the
synapses in the brain according to time interval between presynaptic and
postsynaptic spikes.

We studied the effects of STDP on the topology and spike synchronisation.
Regarding the final topology and depending on the balance between inhibitory and
excitatory couplings, the network can evolve not only to different coupling
strength configurations, but also to different connectivities.

When the strength of the inhibitory connections is of the same order of that of
the excitatory connections, the final topology in the plastic brain exhibits 
the rich-club phenomenon, where neurons that have high degree connectivity 
towards neurons of the same presynaptical group (either excitatory of 
inhibitory) become strongly connected to neurons of the other postsynaptical 
group, i.e., a presynaptical neuron that is highly connected to presynaptical 
excitatory neurons (or inhibitory ones) becomes strongly connected to 
postsynaptical inhibitory (or excitatory ones).

When the strength of the synapses becomes reasonably larger than the strength 
of the excitatory connections, then the final topology has all the features of 
a complex topology, where neurons only sparsely connect to other neurons with a
non-trivial topology.

When noise is introduced in the neural network, we observe that inhibitory 
synapses are always enhanced in the plastic brain. Excitatory synapses can also
be enhanced if the initial network has sufficiently large excitatory synaptic 
strength (no less than about half the value of the  inhibitory synapsis 
strength).

The changes in the synapse strength and the connectivities due to STDP
produce significant alterations in the synchronous states of the neuronal
network. We observe that the synchronous states depend on the balance between
the excitatory and inhibitory intensities. We also find coexistence of
strongly synchronous and weakly synchronous behaviours.


\section*{Acknowledgements}
This work was possible by partial financial support from the following
Brazilian government agencies: CNPq (154705/2016-0, 311467/2014-8), CAPES,
Funda\-\c c\~ao Arauc\'aria, and S\~ao Paulo Research Foundation (processes
FAPESP 2011/19296-1, 2015/07311-7, 2016/16148-5, 2016/23398-8, 2015/50122-0).
Research supported by grant 2015/50122-0 S\~ao Paulo Research Foundation
(FAPESP) and DFG-IRTG 1740/2.



\begin{thebibliography}{}
\bibitem{gerstner02}
W. Gerstner, W. Kistler, Spiking neuron models: Single neurons, populations,
plasticity (Cambridge University Press, Cambridge, 2002)  
\bibitem{sporns05}
O. Sporns, G. Tononi, R. K\"otter. PLoS Comput. Biol. {\bf 1}(4), e42 (2005)
\bibitem{viana14}
R.L. Viana, F.S. Borges, K.C. Iarosz, A.M. Batista, S.R. Lopes, I.L. Caldas.
Commun. Nonlinear Sci. Numer. Simul. {\bf 19}(1), 164 (2014)  
\bibitem{borges15}
F.S. Borges, E.L. Lameu, A.M. Batista, K.C. Iarosz, M.S. Baptista, R.L. Viana.
Physica A {\bf 430}, 236 (2015).
\bibitem{wolfram83}
S. Wolfram. Rev. Mod. Phys. {\bf 55}(3), 601 (1983)
\bibitem{batista12}
C.A.S. Batista, E.L. Lameu, A.M. Batista, S.R. Lopes, T. Pereira, G.
Zamora-L\'opez, J. Kurths, R.L. Viana. Phys. Rev. E {\bf 86}, 016211 (2012)
\bibitem{lameu16a}
E.L. Lameu, F.S. Borges, R.R. Borges, K.C. Iarosz, I.L. Caldas, A.M. Batista,
R.L. Viana, J. Kurths. Chaos {\bf 26}, 043107 (2016)
\bibitem{lameu16b}
E.L. Lameu, F.S. Borges, R.R. Borges, A.M. Batista, M.S. Baptista, R.L. Viana.
Commun. Nonlinear Sci. Numer. Simul. {\bf 34}, 45 (2016)
\bibitem{schappo17}  
M. Girardi-Schappo, G.S. Bortolotto, R.V. Stenzinger, J.J. Gonsalves, M.H.R.
Tragtenberg. PLoS ONE {\bf 12}(3), e0174621 (2017)  
\bibitem{abbott99}
L.F. Abbott. Brain Res. Bull. {\bf 50}, 303 (1999)
\bibitem{batista14}
C.A.S. Batista, R.L. Viana, S.R. Lopes, A.M. Batista. Physica A {\bf 410},
628 (2014)  
\bibitem{baptista10}
M.S. Baptista, F.M. Kakmeni, C. Grebogi. Phys. Rev. E {\bf 82}, 036203 (2010)
\bibitem{lapicque07}
L. Lapicque. J. Physiol. Pathol. Gen. {\bf 9}, 620 (1907)
\bibitem{hodgkin52}
A.L. Hodgkin, A.F. Huxley. J. Physiol. {\bf 117}, 500 (1952)
\bibitem{hindmarsh84}
L.J. Hindmarsh, R.M. Rose. Proc. R. Soc. Lond. B {\bf 221}, 87 (1984)
\bibitem{baptista08}
M.S. Baptista, J. Kurths. Phys. Rev. E {\bf 77}, 026205 (2008)
\bibitem{baptista08plos}
M.S. Baptista, J.X. de Carvalho, M.S. Hussein. PloS ONE {\bf 3}, e3479 (2008) 
\bibitem{cris15}
C.G. Antonopoulos, S. Srivastava, S.S. Pinto, M.S. Baptista. PLoS Comput. Biol.
{\bf 11}, e1004372 (2015) 
\bibitem{uhlhaas09}
P. Uhlhaas, G. Pipa, B. Lima, L. Melloni, S. Neuenschwander, D. Nikoli\'c,
W. Singer. Front. Integr. Neurosci. {\bf 3}, 17 (2009)  
\bibitem{melloni07}
L. Melloni, C. Molina, M. Pena, D. Torres, W. Singer, E. Rodriguez. J. Neurosci.
{\bf 27}(11), 2858 (2007)
\bibitem{borges17}
F.S. Borges, P.R. Protachevicz, E.L. Lameu, R.C. Bo\-netti, K.C. Iarosz,
I.L. Caldas, M.S. Baptista, A.M. Batista. Neural Netw. {\bf 90}, 1 (2017)
\bibitem{fell11}
J. Fell, N. Axmacher. Nat Rev. Neurosci. {\bf 12}, 105 (2011)
\bibitem{rubchinsky12}
L.L. Rubchinsky, C. Park, R.M. Worth. Nonlinear Dyn. {\bf 68}, 329 (2012)
\bibitem{lameu16}
E.L. Lameu, F.S. Borges, R.R. Borges, K.C. Iarosz, I.L. Caldas, A.M. Batista,
R.L. Viana, J. Kurths. Chaos {\bf 26}, 043107 (2016)
\bibitem{benett64}
E.L. Bennett, M.C. Diamond, D. Krech, M.R. Rosenzweig. Science {\bf 146}, 610
(1964)
\bibitem{james90}
W. James, The principles of psychology (Henry Holt and Company, New York, 1890)
\bibitem{lashley23}
K.S. Lashley. Psychol. Bull. {\bf 30}, 237 (1923)
\bibitem{bennett64}
E.L. Bennett, M.C. Diamond, D. Krech, M.R. Rosenzweig. Science {\bf 146}, 610
(1964)  
\bibitem{diamond64}
M.C. Diamond, D. Krech, M.R. Rosenzweig. J. Comp. Neurol. {\bf 123}, 111 (1964)
\bibitem{hebb49}
D.O. Hebb, The organization of behavior (Wiley, New York, 1949)
\bibitem{gerstner12}
W. Gerstner, H. Sprekeler, G. Deco. Science {\bf 338}, 60 (2012)
\bibitem{markram12}
H. Markram, W. Gerstner, P.J. Sjostrom. Front. Synaptic Neurosci. {\bf 4}, 1
(2012)
\bibitem{borges16}
R.R. Borges, F.S. Borges, E.L. Lameu, A.M. Batista, K.C. Iarosz, I.L. Caldas,
R.L. Viana, M.A.F. Sanju\'an. Commun. Nonlinear Sci. Numer. Simul. {\bf 34},
12 (2016)
\bibitem{poo98}
G.-Q. Bi, M.-M. Poo. J. Neurosci. {\bf 18}(24), 10464 (1998)
\bibitem{haas06}
J.S. Haas, T. Nowotny, H.D.I. Abarbanel. J. Neurophysiol. {\bf 96}, 3305 (2006)
\bibitem{alberts02} 
B. Alberts, A. Johnson, J. Lewis, M. Raff, K. Roberts, P. Walter, Molecular
biology of the cell, 4th ed. (Garland Science, New York, 2002)
\bibitem{arbib02}
M.A. Arbib, The handbook of brain theory and neural networks (The MIT Press,
Cambridge 2002)
\bibitem{gouaux05}
E. Gouaux, R. Mackinnon. Science {\bf 310}(5), 344 (2009)
\bibitem{arenas08}
A. Arenas, A. D\'iaz-Guilera, J. Kurths, Y. Moreno, C. Zhou. Phys. Rep.
{\bf 469}, 93 (2008)
\bibitem{deco11}
G. Deco, A. Buehlmann, T. Masquelier, E. Hugues. Front. Hum. Neurosci. {\bf 5},
1 (2011)
\bibitem{popovych13}
 V.O. Popovych, S. Yanchuk, P.A. Tass. Sci. Rep. {\bf 3}, 2926 (2013)
\bibitem{kuramoto84}
Y. Kuramoto, Chemical oscillations, waves, and turbulence (Springer, Berlin,
1984)  
\bibitem{gerstner10}
W. Gerstner. Front. Synaptic Neurosci. {\bf 2}, 1 (2010)
\bibitem{feldman12}
D.E. Feldman. Neuron {\bf 75}, 556 (2012)
\bibitem{bliss73}
T.V. Bliss, T. Lomo. J. Physiol. {\bf 232}, 331 (1973)
\bibitem{song00}
S. Song, K.D. Miller, L.F. Abbott. Nat. Neurosci. {\bf 3}, 919 (2000)
\bibitem{gerstner96}
W. Gerstner, R. Kempter, J.L. van Hemmen. Nature {\bf 383}, 76 (1996)  
\bibitem{markram95}
H. Markram, B. Sakmann. Soc. Neurosci. Abstr. {\bf 21}, 1 (2007)
\bibitem{markram97}
H. Markram, J. L\"ubke, M. Frotscher, B. Sakmann. Science {\bf 275}, 213 (1997)
\bibitem{fregnac10}
Y. Fr\'egnac, M. Pananceau, A. Ren\'e, N. Huguet, O. Marre, M. Levy, D.E.
Schulz. Front. Synaptic Neurosci. {\bf 2}, 73 (2010)
\bibitem{buchanan10}
K.A. Buchanan, J. R. Mellor. Front. Synaptic Neurosci. {\bf 2}, 94 (2010)  
\bibitem{noback05}
C.R. Noback, N.L. Strominger, R.J. Demarest, D.A. Ruggiero, The human nervous
systems: Structure and function, 6th ed. (NJ: Humana Press, Totowa, 2005)
\bibitem{towlson13}
E.K. Towlson, E. V\'ertes, S.E. Anhert, W.R. Schafer, E.T. Bullmore. J. Neurosc.
{\bf 33}, 6380 (2013)
\bibitem{vertes14}
P.E. V\'ertes, A. Alexander-Bloch, E.T. Bullmore. Philos. Trans. R. Soc. Lond.
B Biol. Sci. {\bf 369}, 20130531 (2014)
\end{thebibliography}
\end{document}